\documentclass[aps,prl,reprint,superscriptaddress,showpacs]{revtex4-1}
\usepackage{graphicx}
\usepackage{amssymb}
\usepackage{amsmath}
\usepackage{bm}
\usepackage{verbatim}
\usepackage[bookmarks=false,colorlinks=true,urlcolor=blue,citecolor=blue,linkcolor=blue]{hyperref}
\usepackage[normalem]{ulem}

\newcommand{\GGG}{\mathcal{G}}
\newcommand{\GG}{\mathcal{G}_{0}}
\newcommand{\ex}{\mathrm{ex}}

\renewcommand{\vec}[1]{\mathbf{#1}}
\newcommand{\ikn}{ik_n}

\renewcommand{\vec}[1]{{\bm #1}}

\begin{document}

\title{Self-Organized Topological Superconductivity in a Yu-Shiba-Rusinov Chain}

\author{Michael Schecter}
\affiliation{Center for Quantum Devices, Niels Bohr Institute, University of Copenhagen, 2100 Copenhagen, Denmark}

\author{Karsten Flensberg}
\affiliation{Center for Quantum Devices, Niels Bohr Institute, University of Copenhagen, 2100 Copenhagen, Denmark}

\author{Morten H. Christensen}
\affiliation{Niels Bohr Institute, University of Copenhagen, 2100 Copenhagen, Denmark}

\author{Brian M. Andersen}
\affiliation{Niels Bohr Institute, University of Copenhagen, 2100 Copenhagen, Denmark}

\author{Jens Paaske}
\affiliation{Center for Quantum Devices, Niels Bohr Institute, University of Copenhagen, 2100 Copenhagen, Denmark}

\date{\today}

\begin{abstract}
We study a chain of magnetic moments exchange coupled to a conventional three dimensional superconductor. In the normal state the chain orders into a collinear configuration, while in the superconducting phase we find that  ferromagnetism is unstable to the formation of a magnetic spiral state. Beyond weak exchange coupling the spiral wavevector greatly exceeds the inverse superconducting coherence length as a result of the strong spin-spin interaction mediated through the subgap band of Yu-Shiba-Rusinov states. Moreover, the simple spin-spin exchange description breaks down as the subgap band crosses the Fermi energy, wherein the spiral phase becomes stabilized by the spontaneous opening of a $p-$wave superconducting gap within the band. This leads to the possibility of electron-driven topological superconductivity with Majorana boundary modes using magnetic atoms on superconducting surfaces.
\end{abstract}

\pacs{75.30.Hx, 75.75.-c, 74.20.Mn, 03.67.Lx}

\maketitle

The prospect of performing topological quantum computation~\cite{kitaev} has stimulated intense investigations into condensed-matter systems harboring Majorana bound states~\cite{alicea,beenakker,flensberg-leijnse,fu-kane1,fu-kane2,lutchyn,oreg,exp1,exp2,exp3,exp4,exp5}. One potential platform involves magnetic atoms arranged in a regular lattice on an $s-$wave superconducting substrate~\cite{yazdani1}. This system has received a renewed interest due to recent scanning tunneling microscopy (STM) data possibly supporting the existence of Majorana bound states in a self-assembled one-dimensional (1d) array of atomically-spaced Fe atoms on the surface of superconducting Pb~\cite{yazdani2,berlin,basel}.

\begin{figure}[b!]
\centering
\includegraphics[width=1.\columnwidth]{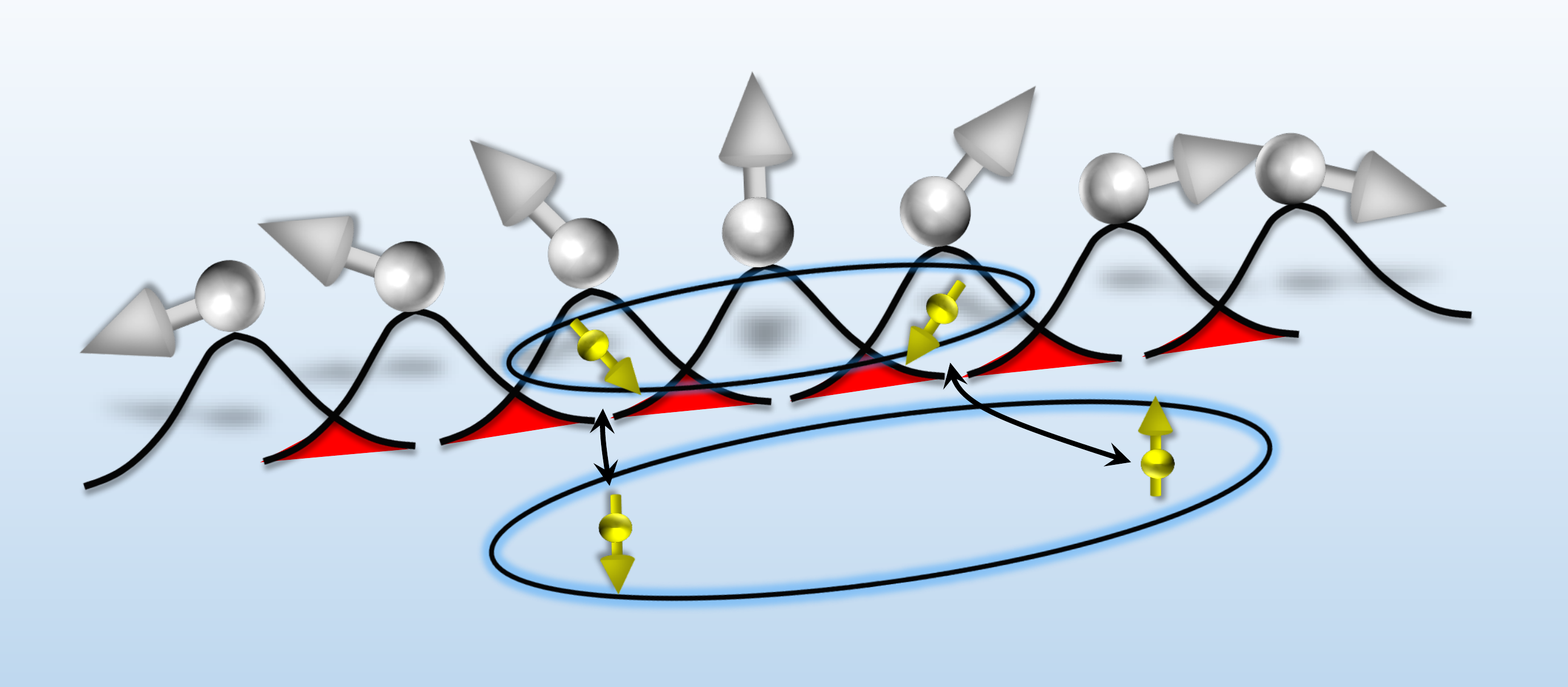}
\caption{(Color online) Magnetic atoms on the surface of a superconductor induce local spin-polarized, subgap YSR states (black curves). YSR wavefunction overlap (shaded red) leads to hopping and $p-$wave pairing due to hybridization with singlet Cooper pairs in the host. This hybridization can drive the spin chain into a spiral state that, in turn, induces spontaneous topological superconductivity within the YSR chain.}
\label{fig:schematic}
\vspace{-0.2 in}
\end{figure}

We focus on the case of an STM-assembled magnetic atom chain whose spacing is several substrate lattice sites \cite{hamburg}, where the overlap of atomic wavefunctions is negligible. In that case each atom acts as an isolated magnetic moment giving rise to localized, sub-gap Yu-Shiba-Rusinov (YSR) states~\cite{yu,shiba,rusinov,balatsky,yazdani4,yazdani5} within the superconductor. Overlap between YSR states leads to the formation of an effectively spinless sub-gap band that can undergo a topological superconducting transition controlled by the magnetic ordering of the magnetic atoms (spin chain)~\cite{pientka1,pientka2}, see Fig.~\ref{fig:schematic}. On the other hand, the ordering of the spin chain is dictated by the indirect exchange mediated by electrons in the superconductor, allowing the possibility of an electron-driven ``self-organized" topological superconducting phase.

Although many important aspects of spin chains on superconductors have been theoretically studied by several authors~\cite{pientka3,pientka1,pientka2,loss,braunecker-simon,vazifeh-franz,sarma1,heimes1,heimes2,brydon-sau,reis-franz} (see also \cite{schecter,choy, martin,kjaergaard}), a generic conceptual account of the collective ordering mechanism of the magnetic and electronic degrees of freedom in a bulk (${\rm d}>1$) superconductor remains elusive (the 1d case is special due to $2k_F$ nesting and was studied in \cite{loss,braunecker-simon,vazifeh-franz}, see also \cite{schecter}).  

\begin{figure*}[t!]
\centering
\includegraphics[width=0.95\columnwidth]{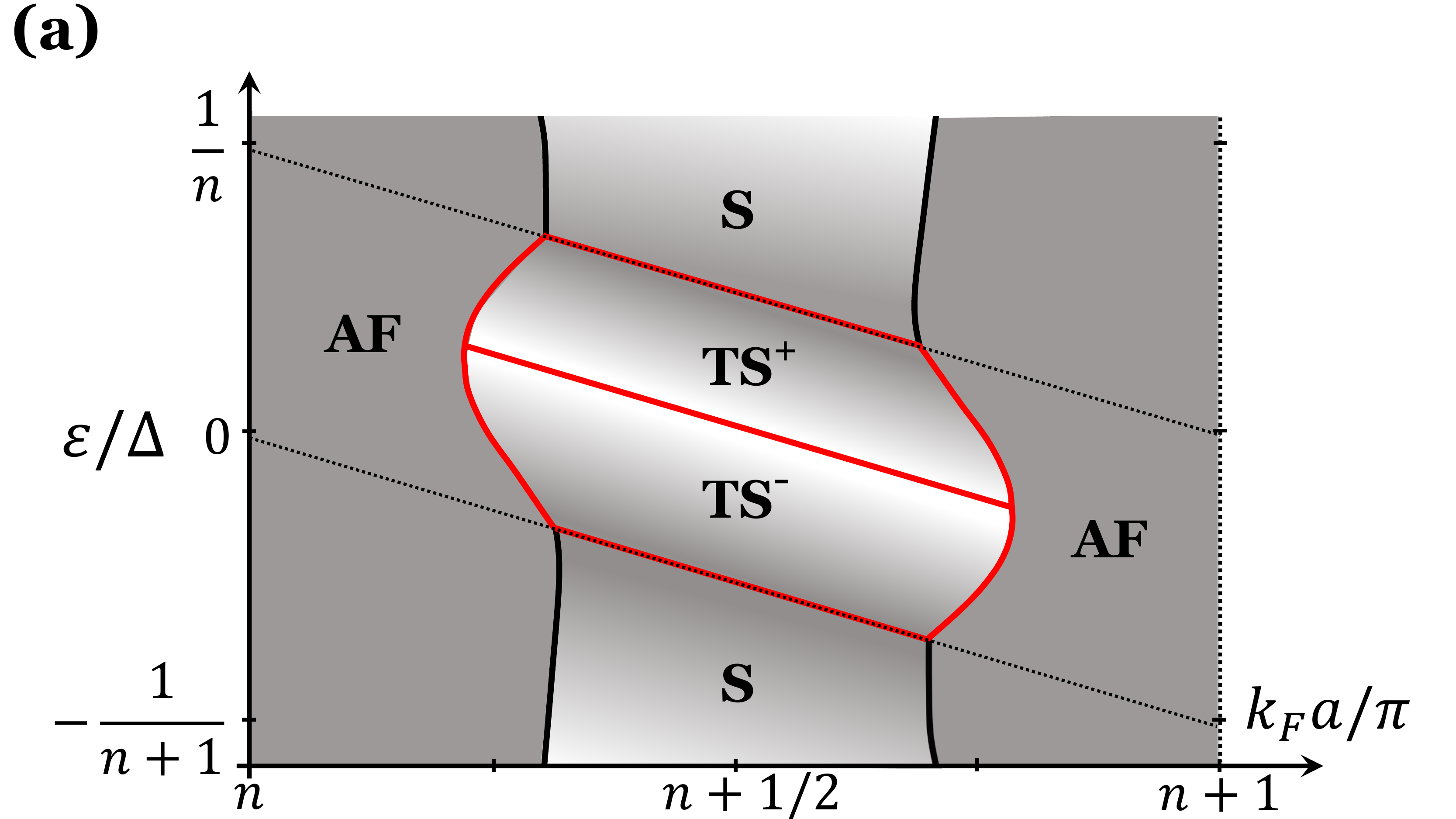}\includegraphics[width=0.95\columnwidth]{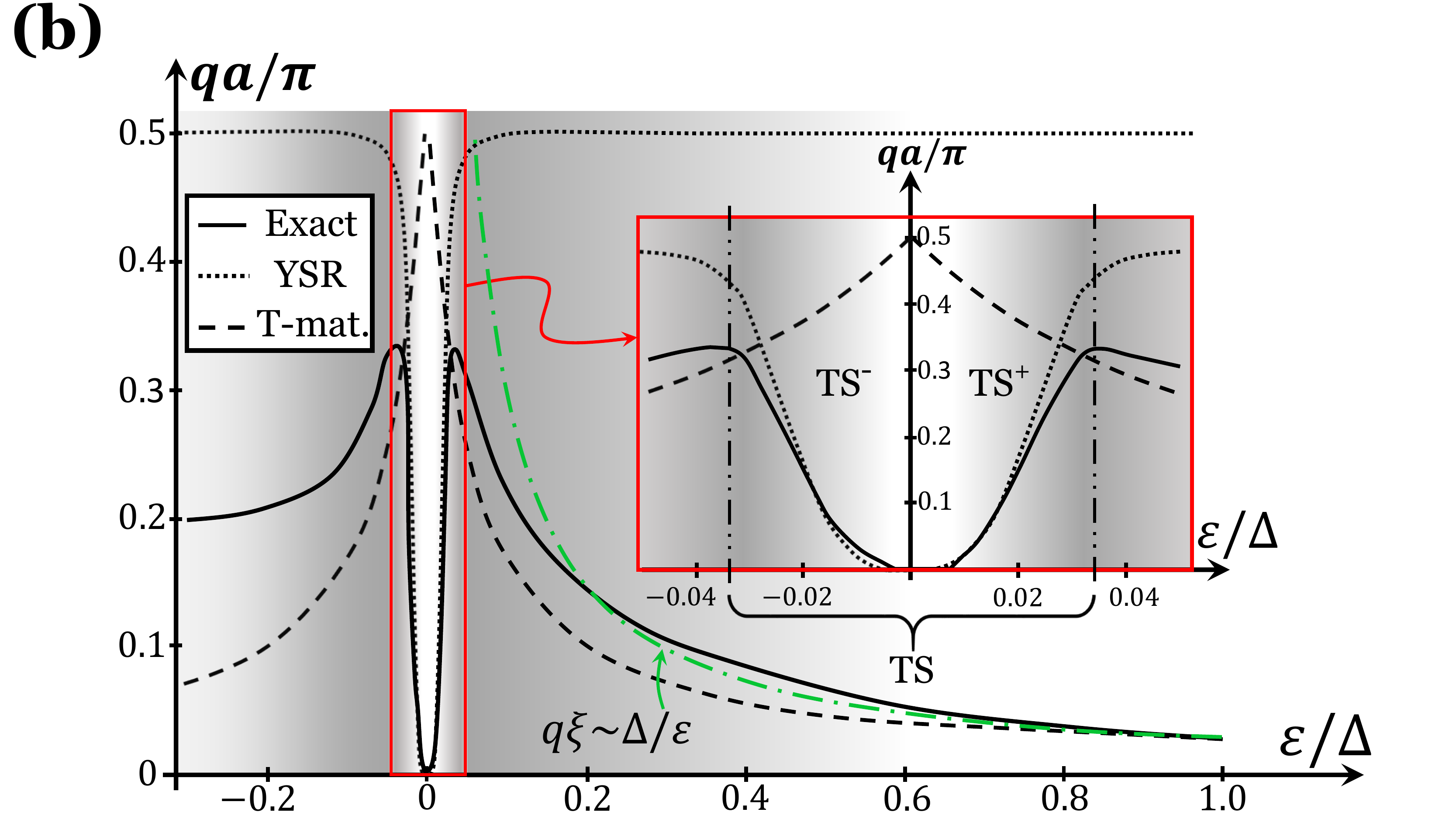}
\caption{(Color online)  \textbf{(a)} Exact groundstate phase diagram of Eq.~(\ref{eq:H}) as a function of lattice spacing and YSR energy $\varepsilon$ (Eq.~\ref{eq:epsilondef}) for integer $n$ ($n=15$ shown) and $k_F\xi=2000$. Outside the range $n+1/4\lesssim k_Fa/\pi\lesssim n+3/4$ the spin chain is an antiferromagnet AF with $q=\pi/a$, while inside this range it is a spiral (S) whose wavevector is shown schematically in gray-scale and plotted in panel \textbf{(b)} along the line $k_Fa=15.5\pi$. The topologically nontrivial superconducting $(\mathrm{TS}^{\pm})$ regions are bounded by the red curves, across which either a first order transition to the AF state or a continuous transition into the trivial S state occurs. The latter transition occurs approximately along the dotted lines: $\varepsilon^{\pm}/\Delta=-(k_Fa\,\,\mathrm{mod}\,\mp\pi)/k_Fa$. The system is gapless along the ferromagnetic line $\varepsilon/\Delta=\left(\pi/2-(k_Fa\,\,\mathrm{mod}\,\pi)\right)/k_Fa$ separating the topologically distinct $\mathrm{TS}^{\pm}$ regions with topological invariants $\pm 1$. \textbf{(b)} Ground-state wavevector $q$ of the magnetic spiral plotted as a function of $\varepsilon$. The solid line corresponds to the exact solution of Eq.~(\ref{eq:H}), the long-dashed line to the $T-$matrix approximation of Eq.~(\ref{eq:J-ex}) and the dotted line to the YSR approximation of Eq.~(\ref{eq:F-YSR}). The inset shows the magnetic order near the topological superconducting (TS$^\pm$) phases bounded by the vertical dash-dotted lines.}
\label{fig:phase-diagram}
\vspace{-0.2 in}
\end{figure*}

In this paper, we establish the existence of spiral magnetic order and self-organized topological superconductivity in the case of a magnetic adatom chain on a 3d substrate using an exactly solvable minimal model. We provide the conceptual framework to describe the ordering mechanism in a general setting, which we find to be entirely distinct from the $2k_F$ mechanism that can occur for a 1d substrate~\cite{loss,braunecker-simon,vazifeh-franz,schecter}. In particular, we show that ferromagnetism of the spin chain is unstable to the formation of a spiral state due to the presence of singlet superconductivity in the host. The corresponding spiral wavevector exhibits a nonmonotonic dependence on the spin-lattice exchange coupling to the superconductor. The wavevector grows rapidly as the exchange increases from zero due to a strong indirect spin interaction mediated by YSR states, and decreases after the YSR band crosses the Fermi energy due to the double-exchange mechanism. In the latter regime the YSR chain exhibits self-sustained $p-$wave superconductivity with the possibility of Majorana boundary modes, depending on the lattice spacing. Our main results are summarized in Fig.~\ref{fig:phase-diagram}, with details related to various technical calculations delegated to the Supplemental Material \cite{supp-mat}.

A system of magnetic atoms exchange coupled to a 3d superconductor may be described by the following Bogoliubov-de Gennes Hamiltonian 

\begin{equation}
\label{eq:H}
H=\frac{1}{2}\sum_\mathbf{k}\Psi_\mathbf{k}^\dagger\left(\xi_\mathbf{k}\tau_z+\Delta\tau_x\right)\Psi_\mathbf{k}
+\frac{1}{2}J\int_\mathbf{r}\Psi_\mathbf{r}^\dagger \vec{S}_\mathbf{r}\cdot\boldsymbol{\sigma}\Psi_\mathbf{r},
\end{equation}
where $\xi(\mathbf{k})=\frac{\mathbf{k}^2-k_F^2}{2m}$ ($\mathbf{k}$, $\mathbf{r}$ and $m$ are the electron momentum, position and mass, respectively, while $k_F$ is the Fermi momentum), $\Psi=(\psi_\uparrow,\psi_\downarrow,\psi_\downarrow^\dagger,-\psi_\uparrow^\dagger)^\mathrm{T}$ is the 4-component Nambu spinor written in terms of electron annihilation (creation) operators $\psi_\sigma\,(\psi^\dagger_\sigma)$ with spin projection $\sigma$. Here $\sigma_i$ and $\tau_i$ are, respectively, Pauli matrices acting in the spin and particle-hole spaces while $\vec{S}_{\vec r}=\sum_j\delta(\vec r-\vec{r}_j)\vec{S}_j$ is the sum of spins $j$ at sites $\vec{r}_j$ with lattice spacing $a$ and $J$ is the exchange interaction constant. 

We provide the exact phase diagram of Eq.~(\ref{eq:H}) within the classical spin approximation \cite{shiba} assuming a planar spiral ansatz, $\vec S_i\cdot\vec S_j=S^2\mathrm{cos}\,Qa(i-j)$, where the spiral wavevector $Q$ is treated as a variational parameter determined by minimizing the total electronic energy of the system, $Q_{\rm min}=q$. The main results of this procedure are shown in Fig.~\ref{fig:phase-diagram}, with technical details provided in the Supplemental Material \cite{supp-mat}. Below, we focus on extracting only the most important and physically relevant ordering mechanisms, which we verify against the exact results.

\emph{Magnetic Atom Order} $-$ The spin-chain magnetic order is easiest to determine in the limit of small exchange coupling, where one may integrate out electrons to second order in $J_{ex}$ to arrive at the effective spin Hamiltonian
\begin{equation}\label{eq:Hspin}
H_{\mathrm{spin}}=\frac{1}{2}\sum_{i\neq j}I(a(i-j))\vec{S}_i\cdot\vec{S}_j.
\end{equation}
The electron-induced spin-spin exchange coupling in Eq.~(\ref{eq:Hspin}) is $(\hbar=1)$ \cite{abrikosov,aristov,galitski-larkin,yao}
\begin{equation}\label{eq:Jr}
I(r)\propto J_{ex}^2e^{-2r/\xi}\left[\frac{v_F}{2\pi r^3}\mathrm{cos}(2k_Fr)+\frac{\Delta}{r^2}\,\mathrm{sin}^2(k_Fr)\right],
\end{equation}
where $v_F$ is the Fermi velocity and $\xi=v_F/\Delta$ is the coherence length of the superconductor. We note that solving for the superconducting order parameter self-consistently is not important in the dilute atom regime ($k_F\gg a^{-1},\,\xi^{-1}$) since the YSR energy and electron propagator between spins are hardly modified \cite{flatte}. We do not consider the dense atom limit $k_Fa<1$, where a collective depletion of $\Delta(r)$ near the chain should be taken into account.

The first term in square brackets in Eq.~\eqref{eq:Jr} is the familiar Rudermann-Kittel-Kasuya-Yosida (RKKY) interaction \cite{rk,kasuya,yosida}, while the second, strictly antiferromagnetic, term arises from the anomalous component of the electronic spin susceptibility and disfavors the pair-breaking effect of a polarized exchange field. The spin-chain groundstate energy is given by the minimum Fourier component of the exchange interaction, $E_0=I_{q}$. 

In the normal state ($\Delta=0$), one finds from Eq.~(\ref{eq:Jr}) a ferromagnet ($q=0$) in the range  $n+1/4<k_Fa/\pi< n+3/4$ with integer $n$ and an antiferromagnet  ($q=\pi/a$) otherwise. We emphasize that the $q=2k_F$ spiral phase that may arise when the electron gas is also one-dimensional \cite{loss,braunecker-simon,vazifeh-franz,schecter} does not occur here; the spin chain is always in a collinear state. However, superconducting correlations in a 3d electron gas generally make ferromagnetism unstable towards spiral formation due to the long-range superconducting correction in Eq.~(\ref{eq:Jr}) (in a 1d electron gas the $q=2k_F$ order is hardly modified by superconductivity~\cite{loss,braunecker-simon,vazifeh-franz}). Indeed, for $\Delta\neq0$ the minimum Fourier component, $I_q\sim q^2-|q|/\xi$, is shifted to the wavevector $q\sim\xi^{-1}$. This magnetic instability is similar to the Anderson-Suhl transition in 2d and 3d spin lattices \cite{anderson-suhl,aristov} and results from the compromise between the shorter-range ferromagnetic RKKY interaction and the longer-range antiferromagnetic interaction.

The spiral formation is stable and strongly reinforced by the inclusion of higher order terms beyond the Born approximation in Eq.~(\ref{eq:Jr}). The crucial effect responsible for this behavior is the formation and overlap of the subgap YSR bound states induced by the magnetic atoms. Formally, these subgap states are encoded in the single magnetic atom $T-$matrices
\begin{align}\label{eq:T-mat}
T_j(\omega)=J\vec{S}_j\cdot\vec\sigma\left(1+\frac{\pi J \nu_{F}}{2}\frac{\omega\tau_{0}+\Delta\tau_{x}}{\sqrt{\Delta^{2}-\omega^2}}\vec{S}_j\cdot\vec\sigma\right)^{-1},
\end{align}
where $\nu_{F}$ is the normal state density of states at the Fermi energy. The presence of the bound state is signified by the subgap pole of $\mathrm{det}\,T_j(\omega)$ on the real energy axis $\omega=\pm \varepsilon$, 
\begin{equation}\label{eq:epsilondef}
\varepsilon=\Delta\frac{1-(\pi J S\nu_{F}/2)^2}{1+(\pi J S\nu_{F}/2)^2}.
\end{equation}
To capture the modification of Eq.~(\ref{eq:Jr}) due to the presence of the YSR bound states, one may replace the bare magnetic atom potential with its $T-$matrix, Eq.~(\ref{eq:T-mat}). Computing the exchange coupling to second order in $T$ leads to \cite{supp-mat} 
\begin{eqnarray}
\label{eq:J-ex}
I(r)&=&\left(1-\frac{\varepsilon^2}{\Delta^2}\right)
\frac{e^{-2r/\xi}}{2(k_{F}r)^2}
\left[\frac{v_{F}}{2\pi r}\mathrm{cos}(2k_{F}r)\right.\\
\nonumber&&\left.+\frac{\Delta^{2}\mathrm{cos}^{2}(k_{F}r)}{2|\varepsilon|}
+\frac{|\varepsilon|}{4}\left[1-3\mathrm{cos}(2k_{F}r)\right]\right],\label{eq:F-2}
\end{eqnarray}
where $J$ is expressed through $\varepsilon$ using Eq.~(\ref{eq:epsilondef}). While in the limit of small $J$, i.e. $\varepsilon\to\Delta$, we recover the perturbative result~\eqref{eq:Jr}, we find for small $\varepsilon$ an enhanced antiferromagnetic component $\sim1/|\varepsilon|$ due to the exchange coupling mediated via the YSR states \cite{yao}. This effect strongly increases the spiral wavevector of the spin chain. Indeed, in the range $n+1/4\lesssim k_Fa/\pi\lesssim n+3/4$ for integer $n$, one finds $I_q\sim q^2-|q|\xi^{-1}\Delta/|\varepsilon|$, with the groundstate wavevector $q\sim\xi^{-1}\Delta/|\varepsilon|$ rapidly growing as $\varepsilon$ approaches zero. This behavior is consistent with the exact numerical result, shown in Fig.~\ref{fig:phase-diagram}b, as long as $|\varepsilon|/\Delta$ is not too small. The apparent divergence of Eq.~\eqref{eq:J-ex} as $|\varepsilon|\to0$ signifies its inapplicability for small enough $\varepsilon$, since the effective exchange energy cannot physically exceed the splitting between YSR energies caused by their  finite  overlap.

This suggests that below a characteristic value of $\varepsilon$, a crossover (for two spins) to a new regime occurs. To see this we follow \cite{yao} and consider two YSR states hybridizing with a Cooper pair from the superconducting condensate. This process can be described by the matrix element $\mathcal{U}(r)=\Delta e^{-r/\xi}\mathrm{cos}(k_Fr)/(k_Fr)$ representing the overlap between YSR states, see Fig.~\ref{fig:schematic}. The energy difference between the ground state and the excited state with YSR states occupied/unoccupied is $2|\varepsilon|$. As long as $\mathcal{U}<2|\varepsilon|$, the change of the groundstate energy can be computed perturbatively, $\mathcal{U}^2/2|\varepsilon|$, leading to the second term in the brackets of Eq.~(\ref{eq:J-ex}). Thus, for the spin chain, Eq.~\eqref{eq:J-ex} is valid down to the scale $|\varepsilon|\sim \mathcal{U}(a)\sim\Delta/k_Fa$ (assuming $\xi\gg a$), at which point $q\sim k_Fa/\xi$. When the Fermi energy crosses the YSR band $|\varepsilon|\lesssim\mathcal{U}(a)\sim\Delta/k_Fa$, the perturbation theory in $\mathcal{U}/|\varepsilon|$ fails since Cooper pairs in the bulk strongly hybridize with the YSR states. This signals a fundamental change of the groundstate, and corresponds to a $p-$wave superconducting phase transition in the YSR chain which harbors Majorana boundary states (provided spiral order remains stable).

\begin{figure}[t!]
\centering
\includegraphics[width=.85\columnwidth]{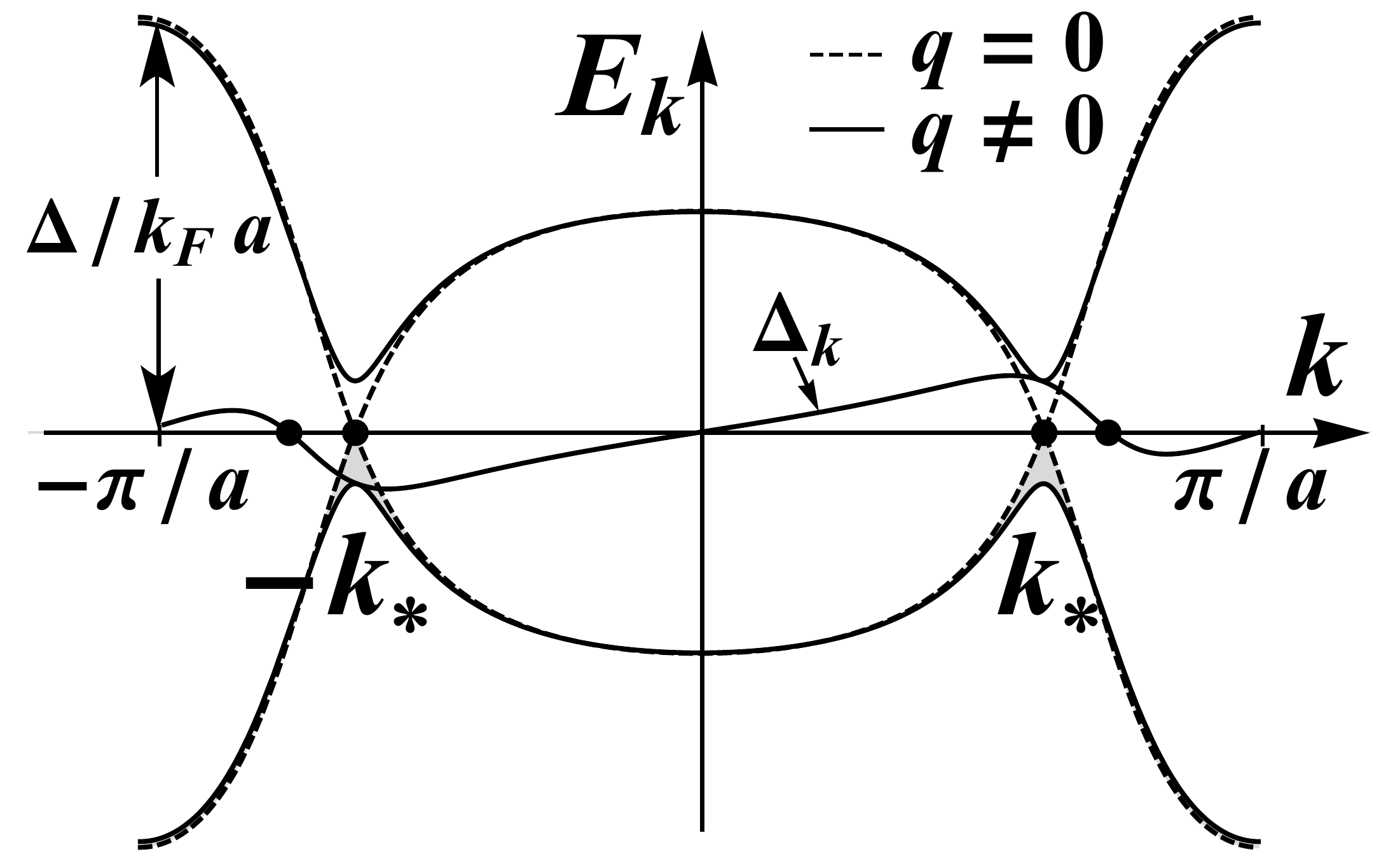}
\caption{Spectrum $E_{k}$ of the YSR band, Eq.~(\ref{eq:F-YSR})
For $q\neq0$
a spectral gap opens and the system gains the condensation energy, corresponding to the gray area between the full and dashed lines for momenta near $\pm k_*$. This occurs only if the pairing at these points is non-zero: $\Delta_{k_*}\neq0$. When the zeroes of $h_k$ and $\Delta_k$ (depicted by dots) coincide, the energy (\ref{eq:F-YSR}) is minimized at $q=0$, corresponding to a gapless electron system and a ferromagnetic spin chain. This phase occurs along lines in the phase diagram given by Eq.~(\ref{eq:f-line}).}
\label{fig:BdG}
\vspace{-0.2 in}
\end{figure}

We now wish to show that in the regime of strong hybridization (where the effective spin model Eqs.~\eqref{eq:Hspin}, \eqref{eq:J-ex} is inapplicable) ferromagnetism remains unstable to spiral formation (i.e. the topological superconductivity is stable). The complication is that a new competing tendency arises once the Fermi energy lies in the YSR band: the gain of kinetic energy associated with electrons hopping along the YSR chain is maximized for a ferromagnetic arrangement of the spin chain since this maximizes the YSR bandwidth, analogous to the double-exchange mechanism. However, the presence of a spiral leads to a superconducting gap in the YSR band, leading to a concomitant gain of the corresponding condensation energy, depicted in Fig.~\ref{fig:BdG}. At small $q$, it is easy to see that the energy gain from the opening of the superconducting gap $\delta E_{\rm gap}\propto -q^2( \partial_q\Delta_{k_*})^2{\rm ln}\frac{1}{qa}$ is always larger than the loss of kinetic (or RKKY) energy $\delta E_{\rm kin}\propto q^2$, where the coefficient of the logarithm is proportional to  the square of the YSR gap at the YSR Fermi points $\pm k_*$ (see Fig.~\ref{fig:BdG}) expanded for small $q$, $\Delta_{k_*}\approx q \partial_q\Delta_{k_*}$ ($\Delta_k=0$ for $q= 0$ since Cooper pairs cannot hybridize with a ferromagnetic chain). Qualitatively, we thus find that ferromagnetism is generically unstable to the formation of spiral phase with wavevector $q\propto e^{-1/\lambda}$, where the parameter $\lambda\propto (\partial_q\Delta_{k_*})^2$ entering the exponent depends on $\varepsilon,\,k_Fa$, and is provided below Eq.~\eqref{eq:f-line}. The case $q=0$ can only occur if the gap at the YSR Fermi points, and thus $\lambda$, vanishes for $k_*\neq 0,\pi/a$.

We study this possibility and the quantitative variation of $q$ on $\varepsilon$ by evaluating the YSR contribution to the total energy numerically (the above gap RKKY contribution, first term in the brackets of Eq.~\eqref{eq:J-ex}, only weakly depends on $|\varepsilon|\ll\Delta$ here)
\begin{align}\label{eq:F-YSR}
E_{\mathrm{YSR}}=-\frac{1}{2}\sum_{k}E_{k};\,\,\,\,\,\,E_k=\sqrt{(h_k-\varepsilon)^2+\Delta_k^2}.
\end{align} 
Here the hopping $h_{k}=h_{-k}$ and $p-$wave pairing $\Delta_{k}=-\Delta_{-k}$ energies are the same as those derived in Ref.~\cite{pientka1}, up to the energy shift $\varepsilon$ which enters Eq.~(\ref{eq:F-YSR}) only as an effective chemical potential for the YSR band. The precise forms of $h_k,\,\Delta_k$ are provided in the Supplemental Material \cite{supp-mat}, but are not essential in the following; their salient features are sketched in Fig.~\ref{fig:BdG}. In particular, one sees that $\Delta_k$ indeed may have extra nodes away from $k=0,\pi/a$ due to the long-range nature of the pairing (i.e. higher odd harmonics beyond ${\rm sin}ka$ are relevant).

Generally we find that when $\Delta_{k_*}=h_{k_*}=0$ have simultaneous zeroes double-exchange wins and the system exhibits a gapless electronic spectrum and ferromagnetism in the spin chain. Remarkably, the exact location of these gapless critical lines in the phase diagram of Fig.~\ref{fig:phase-diagram}a can be extracted:
\begin{equation}
\label{eq:f-line}
\varepsilon_c=\frac{\Delta}{k_Fa}\left[\pi/2-(k_Fa\,\,\mathrm{mod}\,\pi)\right].
\end{equation}
In the spiral regions of Fig.~\ref{fig:phase-diagram}, the magnetic order in the close vicinity of Eq.~(\ref{eq:f-line}) is exponentially suppressed $qa\sim e^{-1/\lambda}$ where $\lambda\sim\left(\frac{\varepsilon-\varepsilon_c}{\Delta/k_Fa}\right)^2$. Away from this line $q$ grows to the scale $q\sim k_Fa/\xi$ at $\varepsilon\sim \Delta/k_Fa$, beyond which it crosses over to the $T-$matrix dependence $q\xi\sim\Delta/|\varepsilon|$, as shown in Fig.~\ref{fig:phase-diagram}b for the case $k_Fa=15.5\pi$ ($\varepsilon_c=0$). This leads to the non-monotonic behavior of $q$ on the YSR energy $\varepsilon$, with a peak around $\varepsilon\sim\Delta/k_Fa$ of order $q_{\mathrm{max}}\sim k_Fa/\xi$. The magnetic order determined from minimizing the electronic energy of Eq.~(\ref{eq:H}), shown in Fig.~\ref{fig:phase-diagram}b, indeed exhibits this crossover behavior and is well-described by the $T-$matrix approximation when $|\varepsilon|>\Delta/k_Fa$ and the YSR approximation when $|\varepsilon|<\Delta/k_Fa$.  


We now briefly discuss the topological properties of the model in Eq.~(\ref{eq:H}), delegating a more detailed account to the Supplemental Material. As discussed in Refs.~\cite{tewari1,tewari2,ojanen1,heimes1,heimes2,ojanen2}, the topological class of the YSR chain in the planar spiral configuration is BDI due to hidden time-reversal and chiral symmetries. Different topological phases are therefore distinguished by a $\mathbb{Z}$ topological invariant (winding number), which can only change when the gap in the YSR spectrum closes. This occurs when the YSR band crosses the Fermi energy or when the system is tuned across the gapless ferromagnetic line Eq.~(\ref{eq:f-line}), leading to the topologically non-trivial and distinct TS$^\pm$ phases with winding numbers $\pm1$ (for a fixed rotation axis of the spiral), shown in Fig.~\ref{fig:phase-diagram}a. As a result, the boundary between TS$^+$ and TS$^-$ systems will host a pair of zero-energy Majorana bound states protected by chiral symmetry, whereas the boundary of either system with vacuum will only host a single Majorana bound state.  The presence of two Majorana bound states also occurs for a single system in either of the TS$^\pm$ phases if there is a $q\to-q$ domain wall \cite{ojanen1,ojanen2}.

\emph{Experimental Outlook} $-$  The YSR topological gap, $\Delta_{\rm t}$, obtains a broad maximum as a function of magnetic atom spacing (i.e. from one period $n$ of the topologically non-trivial part of the phase diagram to the next, $k_Fa\to k_Fa+\pi$). This is because if the magnetic atoms are spaced too closely ($a\to0$) $\Delta_{\rm t}\propto q_{\rm max}a\Delta/k_Fa\propto \Delta a/\xi$ becomes small, while if the atoms are too dilute ($a\to\infty$) the overall energy scale $\Delta/k_Fa$ is diminished. The maximum thus occurs for an optimal spacing $a_{\rm opt}$ determined by $q_{\rm max}a\sim 1$, giving $k_Fa_{\rm opt}\sim\sqrt{k_F\xi}$. At this spacing, for $k_F\xi=2000$, we find $\Delta_{\rm t}\sim 200{\rm mK}$, assuming $\Delta=10$ K (by comparison for Fe on Pb experiments show $\Delta_{\rm t}\lesssim 1\,{\rm K}$ \cite{yazdani2,berlin}). 

The influence of anisotropy and temperature on the topological phase are expected to be unimportant as long as the associated energy scale remains below $\Delta_{\rm t}$. The magnetic order can also be destroyed by thermal fluctuations, but such effects will be quenched either by finite chain size or anisotropy \cite{loss,braunecker-simon,vazifeh-franz,heimes2}. For the case of single-ion anisotropy, described by the Hamiltonian $H_{\mathrm{D}}=-DS^{z2}_j$, we find that when $D<0$ (easy-plane) the spiral is merely fixed to lie in the $xy$-plane (i.e. the spiral axis lies along $\hat z$), while for $D>0$ the spiral axis lies in the $xy-$plane and is orthogonal to the $\hat z$ axis. For sufficiently strong anisotropy $D\gtrsim \Delta_{\rm t}\sim20\mu$eV we expect the groundstate to become a topologically trivial, collinear magnetic phase. This restricts the self-organized topological phase to easy-plane systems or those with sufficiently weak easy-axis anisotropy.

\acknowledgments{ The authors thank M. S. Rudner for a careful reading of the manuscript. The Center for Quantum Devices is funded by the Danish National Research Foundation. The research was  supported by The Danish Council for Independent Research | Natural Sciences. M. H. C. and B. M. A. acknowledge financial support from a Lundbeckfond fellowship (grant A9318).}

\pagebreak
\onecolumngrid
\begin{center}
\textbf{\large Self-Organized Topological Superconductivity in a Yu-Shiba-Rusinov Chain}
\newline
\newline
\textbf{\large  Supplemental Material}
\end{center}

\setcounter{equation}{0}
\setcounter{figure}{0}
\setcounter{table}{0}
\setcounter{page}{1}
\makeatletter
\renewcommand{\theequation}{S\arabic{equation}}
\renewcommand{\thefigure}{S\arabic{figure}}
\renewcommand{\bibnumfmt}[1]{[S#1]}
\renewcommand{\citenumfont}[1]{S#1}

In this Supplemental Material we compute the exact groundstate energy of the BdG Hamiltonian as a function of the spiral wavevector in Sec.~\ref{sec:g-energy} and derive the formula determining the exact subgap YSR spectrum in Sec.~\ref{sec:YSR-spec}, which allows us to classify and determine the exact topological phase boundaries in Sec.~\ref{sec:top}.

\section{Groundstate Energy}
\label{sec:g-energy}

We start from the BdG Hamiltonian (Eq.~(\ref{eq:H}) of the main text) and compute the thermodynamic potential $\Omega=-\frac{1}{\beta}\mathrm{ln}Z$  at inverse temperature $\beta$ using the quantum partition function $Z=\mathrm{Tr}e^{-\beta H}$. This allows us to extrapolate to the groundstate energy in the limit $\beta\to\infty$.  Following Ref.~\cite{bruus} (see p. 254), we introduce the real parameter $\lambda$ multiplied onto the exchange part of the BdG Hamiltonian~\eqref{eq:H},
\begin{align}\label{eq:Hlam}
H_{\lambda}=
\frac{1}{2}\sum_\mathbf{p}\Psi_\mathbf{p}^\dagger\left(\xi_\mathbf{p}\tau_z+\Delta\tau_x\right)\Psi_\mathbf{p}
+\frac{1}{2}\lambda\sum_{j}\Psi_{j}^\dagger{\rm J}_{j}\Psi_{j},
\end{align}
written with $4\times 4$ Nambu matrices ${\rm J}_{j}=J_\ex({\vec S}_{j}\cdot\boldsymbol{\sigma})\tau_{0}$. From this definition, the change of thermodynamic potential due to the spin lattice can be found as an integral of $d\Omega/d\lambda$ over $\lambda$:
\begin{align}
\Delta\Omega =\Omega(1)-\Omega(0)
=\int_{0}^{1}d\lambda\sum_{j}\langle \frac{1}{2}\Psi_{j}^\dagger{\rm J}_{j}\Psi_{j}\rangle_{\lambda}=\frac{1}{2\beta}\int_{0}^{1}d\lambda{\rm Tr}\left[{\mathbb J}\,{\mathbb G}^{\lambda}\right].\label{eq:trvg}
\end{align}
Here we use the matrix notation ${\mathbb J}_{ij}={\rm J}_{i}\delta_{ij}$ and ${\mathbb G}^{\lambda}_{ij}={\mathcal G}^{\lambda}(\ikn;r_{i},r_{j})$, with the Nambu matrix structure being implied. The trace involves a sum over spin lattice positions, $\{i\}$, Nambu indices and Matsubara frequencies. Notice that the Green function ${\mathcal G}^{\lambda}$ is defined with averaging, $\langle\ldots\rangle_{\lambda}$, with respect to $H_{\lambda}$.

In terms of the bare Green function, ${\mathbb G}^{0}$, the dressed Green function matrix is obtained by summing the Dyson series
\begin{equation}\label{eq:dyson}
{\mathbb G}^{\lambda}={\mathbb G}^{0}+{\mathbb G}^{0}\lambda{\mathbb J}{\mathbb G}^{0}+{\mathbb G}^{0}\lambda{\mathbb J}{\mathbb G}^{0}\lambda{\mathbb J}{\mathbb G}^{0}+\ldots,
\end{equation}
We note that the 4-spinor notation, $\Psi=(\psi_\uparrow,\psi_\downarrow,\psi_\downarrow^\dagger,-\psi_\uparrow^\dagger)^\mathrm{T},$ exhibits the usual particle-hole redundancy since $\Psi^{\dagger}(x)=\tau^{y}\sigma^{y}\Psi(x)$. The energy overcounting is avoided by the factor of 1/2 multiplying the Hamiltonian (\ref{eq:Hlam}). This factor of 1/2 also compensates a factor of 2 appearing at each vertex in the Dyson series (\ref{eq:dyson}) due to the fact that the perturbation series for the Green function involves both anomalous  $\langle T_{\tau}\Psi(\tau)\Psi(0)\rangle_{0}$ as well as  normal $\langle T_{\tau}\Psi(\tau)\Psi^{\dagger}(0)\rangle_{0}$  Nambu contractions. Since, however, the two are related by multiplication with $\tau^{y}\sigma^{y}$ from the right, their contributions are equivalent and give rise to a compensating factor of 2 at each vertex $\frac{1}{2}{\rm J}_{i}$. As a result, one arrives at the Dyson series~\eqref{eq:dyson} containing only Nambu normal Green functions.

Inserting~\eqref{eq:dyson} into ~\eqref{eq:trvg}, the $\lambda$-integration can be carried out order by order replacing $\lambda^{n-1}$ by $1/n$, and the infinite series can be recognized as a logarithm:
\begin{align}
\Delta\Omega=&-\frac{1}{2\beta}{\rm Tr}\ln\left[1-{\mathbb J}{\mathbb G}^{(0)}\right].\label{eq:dE}
\end{align}
Since we are interested in the magnetic ordering arising from the induced interaction between different spins, we focus on the part of the energy shift which depends on the spin lattice spacing, $a$. The only $a$-independent terms in the Dyson equation are those which only involve repeated scattering from a single magnetic atom. To facilitate this separation we split the bare Green function matrix into a sum of local ($\propto\delta_{ij}$), and non-local ($\propto 1-\delta_{ij}$) terms:
\begin{align}
{\mathbb G}^{0}={\mathbb G}^{0}_{loc}+\tilde{\mathbb G}^{0}.
\end{align}
The $a$ dependent part of~\eqref{eq:dE} is therefore found by subtracting the local part, i.e.
\begin{align}\label{eq:F}
\Delta\Omega_{a}=-\frac{1}{2\beta}\left({\rm Tr}\ln\left[1-{\mathbb J}{\mathbb G}^{0}\right]-
{\rm Tr}\ln\left[1-{\mathbb J}{\mathbb G}_{loc}^{0}\right]\right)
=-\frac{1}{2\beta}{\rm Tr}\ln\left[1-{\mathbb T}\tilde{\mathbb G}^{0}\right],
\end{align}
where the local $T$-matrix has been introduced as
\begin{align}
{\mathbb T}=\left(1-{\mathbb J}{\mathbb G}_{loc}^{0}\right)^{-1}{\mathbb J}.
\end{align}

To investigate the magnetic structure, we use a variational approach in which we minimize the electronic ground state energy with respect to the magnetic ordering profile. We consider a planar spiral order ansatz whose spiral axis of rotation is defined along $\hat z$
\begin{align}\label{eq:spans}
\vec{S}_\mathbf{r}=S\sum_j\delta(x-ja)\delta(y)\delta(z)\,(\cos qaj,\sin qaj,0),
\end{align}
which interpolates between ferromagnetic (F) $(q=0)$ and anti-ferromagnetic (AF) $(q=\pm\pi/a)$ order \cite{Note2}. Within the spiral ansatz, the $j$ dependence of $T_{j}$ can be conveniently
written as $T_{j}=e^{-i\frac{1}{2}qaj\sigma_{z}}T e^{i\frac{1}{2}qaj\sigma_{z}}$,
where $T=J\sigma_{x}(1-\GG(\ikn,0)J\sigma_{x})^{-1}$. Since $\GG\propto\sigma_{0}$, the Green function commutes with $e^{i\frac{1}{2}qaj\sigma_{z}}$, and we may rewrite Eq.~(\ref{eq:F}) as
\begin{equation}
\Delta\Omega_a=-\frac{1}{2\beta}\mathrm{Tr\,ln}\mathbb M,\,\,\,\,\,\,\,\mathbb{M}_{ij}=\delta_{ij}-(1-\delta_{ij})\GG\left(\ikn,a(i-j)\right)
e^{i\frac{1}{2}qa(i-j)\sigma_{z}}T.\label{eq:M}
\end{equation}
The local Green function and the $T$-matrix are given by
\begin{align}
\GG(\ikn,0)&=-\frac{\pi\nu_{F}}{2}\frac{\ikn\tau_{0}+\Delta\tau_{x}}{\sqrt{k_{n}^{2}+\Delta^{2}}}\sigma_{0},\label{eq:Gloc}\\
T(\ikn)&=J\sigma_x\left(1+\frac{\pi J\nu_{F}}{2}\frac{\ikn\tau_{0}+\Delta\tau_{x}}{\sqrt{k_{n}^{2}+\Delta^{2}}}\sigma_x\right)^{-1}.
\end{align}
As noted in the main text the determinant of $T$ exhibits a pole on the real energy axis $\ikn\to\varepsilon$ where $\varepsilon$ is defined in Eq.~\eqref{eq:epsilondef}. For $r\gg v_{F}/\omega_{D}$, with Debye frequency $\omega_{D}$ representing the high-frequency cut-off in the BCS Green function, the non-local Green function takes the form
\begin{eqnarray}
\label{eq:G-r}
\GG(\ikn,r)=-\frac{\pi \nu_{F}}{2}\frac{e^{-r\sqrt{k_{n}^{2}+\Delta^{2}}/v_F}}{k_Fr}
\left[\frac{\ikn\tau_{0}+\Delta\tau_{x}}{\sqrt{k_{n}^{2}+\Delta^{2}}}\mathrm{sin}(k_{F}r)+\tau_{z}\mathrm{cos}(k_{F}r)\right],
\end{eqnarray}
which applies to propagation along the spin chain, provided $k_{F}a\gg\varepsilon_{F}/\omega_{D}$.

The matrix $\mathbb{M}$ in Eq.~(\ref{eq:M}) exhibits translational invariance, allowing it to be diagonalized using a set of Bloch states labeled by quasi-momentum $|k|<\pi/a$. In this basis, the operator trace over spin lattice positions in Eq.~(\ref{eq:F}) becomes an integral over the Brillouin zone. Focusing on the low-temperature limit $\beta^{-1}\ll\Delta/k_Fa$, we replace the Matsubara summation by a frequency integral to obtain the variational groundstate energy for a chain of $N$ spins
\begin{equation}
\label{eq:E}
E(q) =-\frac{Na}{2}\int\!\frac{d\omega}{2\pi}\frac{dk}{2\pi}\mathrm{Tr}\,
\mathrm{ln}\left[1-\tilde{\GG}(i\omega,k-\frac{1}{2}q\sigma_z) T\right].
\end{equation}
In Eq.~(\ref{eq:E}) we introduced  the Fourier transform of the non-local part of the Green function
\begin{equation}\label{eq:Gk}
\tilde{\GG}(i\omega,k)=\sum_{j\neq0}\GG(i\omega,ja)e^{-ikaj}.
\end{equation}
The remaining trace over the $4\times4$ spin/particle-hole space in Eq.~(\ref{eq:E}) can be performed explicitly. The groundstate magnetic order is characterized by the wavevector $q$ which minimizes Eq.~(\ref{eq:E}). The minimization procedure is carried out numerically, with the main results for the magnetic order of the spin chain displayed in the exact phase diagram of Fig.~\ref{fig:phase-diagram} of the main text.

We note that in the dilute limit $k_Fa\gg1$ we have $\GG T\ll1$ and thus the off-diagonal components of $\mathbb{M}_{ij}$ are small. This is because $\GG$ rapidly falls off as $r^{-(d-1)/2}$ and thereby provides a controlled expansion for $d>1$. We may then approximate the energy by expanding (\ref{eq:F}) to second order in $\GG T$: 
\begin{equation}
E=\frac{1}{4}\sum_{i\neq j}\int\frac{d\omega}{2\pi}\mathrm{Tr}\left[T_{i}\mathcal{G}_{0,ij}T_{j}\mathcal{G}_{0,ji}\right].\label{eq:E-pert}
\end{equation}
Carrying out the remaining trace and frequency integral leads to Eq.~(\ref{eq:J-ex}) of the main text in the limit $r\ll \xi$. For $r>\xi$ the terms in square brackets in Eq.~(\ref{eq:J-ex}) are modified by powers of $r$, but the entire expression $\propto e^{-2r/\xi}$ becomes exponentially suppressed in this limit and we do not consider it further. As discussed in the main text, the perturbative expansion leading to the Eq.~(\ref{eq:E-pert}) fails for $|\varepsilon|<\Delta/k_Fa$ due to strong YSR-Cooper pair hybridization. Formally, the $1/|\varepsilon|$ singularity in Eq.~(\ref{eq:J-ex}) is regularized by keeping all terms under the logarithm in Eq.~(\ref{eq:F}). In that case one must understand the structure of the term under the logarithm in Eq.~(\ref{eq:E}), which encodes the subgap YSR spectrum as well as its topological properties.

\section{YSR subgap spectrum}\label{sec:YSR-spec}
The dispersion relation for the chain of YSR states is determined by having zero determinant under the logarithm in Eq.~(\ref{eq:E}). The sub-gap excitations induced by the spin lattice thus correspond to roots of the equation
\begin{align}
\label{eq:spec}
0=\det\!\left[\tilde{\GG}(\omega,k-q\sigma_z/2)-T^{-1}(\omega)\right],
\end{align}
where we have pulled out a factor of $\mathrm{det}\,T$ for convenience, since it has no sub-gap zeroes. One may also derive Eq.~(\ref{eq:spec}) from the Dyson series for the exact electron Green function (see also Ref.~\cite{ojanen-YSR-S}), given by

\begin{align}
\label{eq:spec1}
\mathbb{G}_{ij}=\mathbb{G}^0_{ij}+\sum_l\mathbb{G}^0_{il}\mathrm{J}_l \mathbb{G}_{lj},
\end{align}
where, within the spiral ansatz of Eq.~(\ref{eq:spans}), we have $\mathrm{J}_l=e^{-i\frac{1}{2}qal\sigma_{z}}J\sigma_x e^{i\frac{1}{2}qal\sigma_{z}}$. Multiplying Eq.~(\ref{eq:spec1}) on the left by $e^{i\frac{1}{2}qai\sigma_z}$ and on the right by $e^{-i\frac{1}{2}qaj\sigma_z}$ we obtain

\begin{align}
\label{eq:spec2}
\mathbb{G}_{ij}e^{i\frac{1}{2}qa(i-j)\sigma_z}&=\mathbb{G}^0_{ij}e^{i\frac{1}{2}qa(i-j)\sigma_z}
+\sum_l\mathbb{G}^0_{il}e^{i\frac{1}{2}qa(i-l)\sigma_z}J\sigma_x \mathbb{G}_{lj}e^{i\frac{1}{2}qa(l-j)\sigma_z}.
\end{align}
In momentum space Eq.~(\ref{eq:spec2}) becomes, in the notation of the main text,

\begin{align}
\label{eq:spec3}
\GGG(ik_n,k-q\sigma_z/2)=\GG(ik_n,k-q\sigma_z/2)
+\GG(ik_n,k-q\sigma_z/2)J\sigma_x\GGG(ik_n,k-q\sigma_z/2),
\end{align}
while the inverse of Eq.~(\ref{eq:spec3}) may be written as
\begin{equation}
\label{eq:h_t}
\GGG^{-1}(\ikn,k-q\sigma_z/2)=\GG^{-1}(\ikn,k-q\sigma_z/2)-J\sigma_{x}.
\end{equation}
Splitting the bare Green function into non-local, and local parts
\begin{align}
\GG(\ikn,k)=\sum_{j}\GG(\ikn,ja)e^{-ikaj}=\tilde\GG(\ikn,k)+\GG(\ikn,0),\label{eq:app-G}
\end{align}
with $\tilde\GG$ defined in Eq.~\eqref{eq:Gk}, the local term and the self-energy may be combined into the $T$-matrix whereby
\begin{equation}\label{eq:invG}
\GGG^{-1}(\ikn,k-q\sigma_z/2)=\GG(\ikn,k-q\sigma_z/2)^{-1}\left(T^{-1}(\ikn)-\tilde{\GG}(\ikn,k-q\sigma_z/2)\right)J\sigma_{x}.
\end{equation}
After analytical continuation, $\ikn\to\omega+i0_{+}$, the poles of $\GGG$ are found as zeroes of the determinant of $\GGG^{-1}$. The subgap solutions of Eq.~(\ref{eq:app-G}) occur solely from the zero determinant of the term in parentheses, leading to the same condition as Eq.~(\ref{eq:spec}), which may be rewritten as
\begin{align}\label{eq:Ginvdet}
0=\det\!\left[
\tilde{B}\sigma_{x}+
\frac{\omega+\Delta\tau_{x}}{\sqrt{\Delta^{2}-\omega^{2}}}(\tilde{\Delta}_{s}+
\tilde{\Delta}_{t}\sigma_{z})+
(\tilde\xi+\tilde\alpha\sigma_z)\tau_z\right],
\end{align}
with $\tilde B=(\pi J\nu_{F}/2)^{-1}=\sqrt{(\varepsilon+\Delta)/(\varepsilon-\Delta)}$ and
\begin{equation}\label{eq:H*}
\begin{aligned}
\tilde\xi&=\mathrm{Re}\,g_{e}(\omega,k),\quad &\tilde\alpha=\mathrm{Re}\,g_{o}(\omega,k),\\
\tilde\Delta_{s}&=1+\mathrm{Im}\,g_{e}(\omega,k), &\tilde\Delta_{t}=\mathrm{Im}\,g_{o}(\omega,k).
\end{aligned}
\end{equation}
Here the functions $g_{e/o}$ are defined as
\begin{align}
\label{eq:g}
g_{e/o}(\omega,k)=-\frac{1}{2k_Fa}\left[\mathrm{ln}(1-e^{-(a/\xi)\sqrt{1-\omega^{2}/\Delta^{2}}+i(k_F+k-q/2)a})+
\mathrm{ln}(1-e^{-(a/\xi)\sqrt{1-\omega^{2}/\Delta^{2}}+i(k_F-k+q/2)a})\pm(k\rightarrow -k)\right].
\end{align}
For a given $k$, Eq.~\eqref{eq:Ginvdet} has four roots placed symmetrically around zero, two of which lie in the continuum above the gap. The sub-gap excitation energy $\omega=E_{k}$ generally takes a rather complicated form, which can be simplified by expanding to second order in $(k_{F}a)^{-1}$ and $\varepsilon/\Delta$, to arrive at:
\begin{equation}\label{eq:spec1}
E_k=\sqrt{(h_k-\varepsilon)^2+\Delta_k^2},
\end{equation}
where $h_k=\Delta \mathrm{Im}\,g_e(0,k)$ and $\Delta_k=-\Delta \mathrm{Re}\,g_o(0,k)$. The identical spectrum Eq.~(\ref{eq:spec1}) was found in Ref.~\cite{pientka1S}. We may now use the more general result (\ref{eq:Ginvdet}) to investigate the topological properties of the YSR chain.


\section{Topological phase transition}\label{sec:top}
One may determine the topological classification by studying the exact inverse Green function at zero frequency, which defines a `topological Hamiltonian' \cite{wang} for the sub-gap electron degrees of freedom that propagate along the spin chain. As we show in Sec.~\ref{sec:top1}, the precise form of the topological Hamiltonian is somewhat arbitrary and thus allows one to choose the most convenient representation. To that end, we define the topological Hamiltonian as the zero-frequency limit of the $4\times 4$ matrix used in Eq.~\eqref{eq:Ginvdet}
\begin{align}\label{eq:Htop}
H_{\mathrm{top}}(k)&=\tilde{\GG}(0,k-\frac{1}{2}q\sigma_z)-T^{-1}(0)\nonumber\\
&=\tilde B\sigma_x+\tilde\xi\tau_z+\tilde\alpha\sigma_z\tau_z
+\tilde\Delta_s\tau_x+\tilde\Delta_t\sigma_z\tau_x,
\end{align}
with coefficients defined as the zero-frequency limit of Eq.~\eqref{eq:H*}. We have written $H_{\mathrm{top}}(k)$ in a suggestive form to make the analogy with the nanowire Hamiltonian of Refs.~\cite{lutchynS,oregS}. In this case, one would interpret $\tilde B$ as an effective Zeeman field, $\tilde\Delta_{s/t}$ as singlet/triplet pairing, $\tilde\xi$ as kinetic energy and $\tilde\alpha$ as spin-orbit coupling.
As discussed in Refs.~\cite{tewari1S,tewari2S,ojanen1S,heimes1S,heimes2S}, the topological class of $H_{\mathrm{top}}$ is BDI due to hidden time-reversal, and chiral symmetries, and is therefore characterized by a $\mathbb{Z}$ topological invariant. This can be shown by noting that, in addition to the usual particle-hole symmetry $\Xi H_{\mathrm{top}}(k)\Xi^{-1}=-H_{\mathrm{top}}(-k)$ where $\Xi=\sigma_y\tau_y\mathcal{K}$ and $\mathcal{K}$ denotes complex conjugation, there is an effective  (anti-unitary)  time-reversal operator $\mathcal{O}=-i\sigma_x\mathcal{K}$ such that $\mathcal{O}H_{\mathrm{top}}(k)\mathcal{O}^{-1}=H_{\mathrm{top}}(-k)$. The operator $\mathcal{O}$ satisfies $\mathcal{O}^2=1$ and is distinct from the canonical time-reversal operator $\Theta=i\sigma_y\mathcal{K}$, $\Theta^2=-1$, which does not commute with $H_{\mathrm{top}}$ in the presence of the Zeeman field $\tilde B$. The existence of $\mathcal{O}$ allows one to define a chiral operator $\mathcal{C}=\mathcal{O}\,\Xi=\sigma_z\tau_y$, $\mathcal{C}^2=1$, that anti-commutes with $H_{\mathrm{top}}$: $\{\mathcal{C},H_{\mathrm{top}}(k)\}=0$. This implies that $H_{\mathrm{top}}(k)$ can be block off-diagonalized in the eigenbasis of $\mathcal{C}$, a representation that may be achieved with the unitary transformation $U=e^{\frac{i\pi}{4}\sigma_z\tau_x}$:
\begin{subequations}
\begin{eqnarray}
\label{eq:AU}&&U^\dagger H_{\mathrm{top}} U=\left(\begin{array}{cc}
0 & A\\
A^{\dagger} & 0
\end{array}\right),\\
\label{eq:A}A&=&i\tilde\alpha+\tilde\Delta_s+\tilde B \sigma_{y}+(\tilde\Delta_t+i\tilde\xi)\sigma_{z}.
\end{eqnarray}
\end{subequations}
The $\mathbb{Z}$ topological invariant measures the degree of the map from the Brillouin zone to the U(1) phase of the complex number $z(k)=\mathrm{det}A/|\mathrm{det}A|$ , $\pi_1(S^1)=\mathbb{Z}$, and is therefore represented by a winding number $W$. From Eq.~(\ref{eq:A}) we find
\begin{equation}
\label{eq:z}
z=\frac{(\pi J\nu_{F}/2)^{2}\left(1-i(g_e-g_o)\right)\left(1+i(g_e^*+g_o^*)\right)-1}{\left|(\pi J\nu_{F}/2)^{2}\left(1-i(g_e-g_o)\right)\left(1+i(g_e^*+g_o^*)\right)-1\right|}
\end{equation}
with
\begin{equation}
\label{eq:W}
W=\frac{1}{2\pi i}\oint\frac{dz}{z}=\frac{1}{2\pi i}\oint dk\partial_k\mathrm{ln}z(k).
\end{equation}
The value of $W$ is quantized and can only change when $z$ becomes ill-defined. This occurs when $\mathrm{det}A=0$, and coincides with the closing of the spectral gap $\mathrm{det}H_{\mathrm{top}}=0$, see Eq.~(\ref{eq:AU}).

Since the function $g_o(0,k)$ vanishes at the time-reversal symmetric points $k=0,\pm\pi/a$, the location of the gap-closing transitions simplify in this case to give
\begin{eqnarray}
\label{eq:gap-close1}
\frac{\varepsilon^\pm}{\Delta}=\frac{\left|1-ig_e^\pm\right|^2-1}{\left|1-ig_e^\pm\right|^2+1},
\end{eqnarray}
where $g_e^\pm=g_e(0,k)$ with $k=0\,\,\mathrm{or}\,\,\pi/a$, respectively. In the limit $k_Fa\gg1$ we find
\begin{eqnarray}
\label{eq:gap-close2}
\frac{\varepsilon^\pm}{\Delta}\approx\mathrm{Im}\,g_e^\pm,
\end{eqnarray}
which is equivalent to the gap closing transition at $k=0,\,\pi/a$ in the YSR band for $k_Fa\gg1$ ($E_k=h_k-\varepsilon=\Delta_k=0$). Furthermore, in the limit $\xi\gg a$ this leads to
\begin{eqnarray}
\label{eq:gap-close3}
\frac{\varepsilon^\pm}{\Delta}\approx-(k_Fa\,\,\mathrm{mod}\,\mp\pi)/k_Fa,
\end{eqnarray}
see Fig.~\ref{fig:phase-diagram}. Other gap closing transitions for $k\neq 0,\pi/a$ are not precluded and do occur. However, their locations in the phase diagram are difficult to determine analytically for arbitrary $k_Fa$. An example of such behavior was discussed in the main text in the limit $k_Fa\gg1$ and was shown to lead to the gapless ferromagnetic line (\ref{eq:f-line}).

\subsection{Topologically Equivalent Hamiltonians}\label{sec:top1}

In this section we show that the definition of the Hamiltonian (\ref{eq:Htop}) leads to the same topological classification as the Hamiltonian $h_{\mathrm{top}}$ defined by the prescription of Ref.~\cite{wang} to be the negative inverse Green function at zero-frequency. According to Eq.~\eqref{eq:invG} this gives
\begin{align}
h_{\mathrm{top}}(k)=\GG(k-\frac{1}{2}q\sigma_z)^{-1}
\left(\tilde{\GG}(k-\frac{1}{2}q\sigma_z)-T^{-1}\right)
J\sigma_{x},
\end{align}
with the (zero) frequency variable suppressed. We may write $h_{\mathrm{top}}=B H_{\mathrm{top}} D=D-B$ (cf. Eq.~(\ref{eq:h_t})), with $D=J\sigma_x$ and $B=\GG(k-\frac{1}{2}q\sigma_z)^{-1}$. Thus, if $B,D$ anti-commute with the chiral operator $\mathcal{C}=\sigma_z\tau_y$, then $C h_{\mathrm{top}}(k)C^{-1}=-h_t(k)$ has the same chiral symmetry as $H_{\mathrm{top}}$. Since one readily sees $C D C^{-1}=-D$, we need only prove that $C B C^{-1}=-B$. Since $B^{-1}=H_{\mathrm{top}}+D^{-1}$ clearly anti-commutes with $C$, one may simply take the inverse of the equation $C B^{-1}C^{-1}=-B^{-1}$ to prove that $B$ also anti-commutes with $C$. Here we have utilized the fact that $B$ is an invertible matrix for every $k$ due to the bare electron spectrum being gapped by superconducting correlations (the zero eigenvalue may occur only in $H_{\mathrm{top}}$).

In the eigenbasis of $\mathcal{C}$, the matrices $D,B$ also become purely off-diagonal, and the off-diagonal element of $h_{\mathrm{top}}(k)$ is therefore of the form $\tilde{B}A\tilde{D}$, where $\tilde B$ and $\tilde D$ are the off-diagonal components of $B$ and $D$, respectively. The winding number $W$ associated with phase of the complex number
\begin{eqnarray}
z^\prime(k)=\frac{\mathrm{det}A}{|\mathrm{det}A|}\frac{\mathrm{det}\tilde B}{|\mathrm{det}\tilde B|}\frac{\mathrm{det}\tilde C}{|\mathrm{det}\tilde C|}
\end{eqnarray}
is unchanged because the determinants of $\tilde B,\tilde C$ never vanish, and hence $W$ may only change when $\mathrm{det}A=0$. Thus, the topological classifications of $h_{\mathrm{top}}$ and $H_{\mathrm{top}}$ are equivalent.

\end{document}